\documentstyle[12pt]{article}
\title{\bf Information entropy as a measure of the quality of a nuclear density
distribution}
\author{G.A. Lalazissis$^{1,2}$, S.E. Massen$^{1}$, C.P. Panos$^{1}$
and S.S Dimitrova$^{3}$\\
\    \\
$^{1}$ Department of Theoretical Physics, 
Aristotle University of Thessaloniki\\
GR 54006 Thesssaloniki, Greece\\ 
$^{2}$ Physikdepartment der Technischen Universit\"at M\"unchen\\
D-85747 Garching bei M\"unchen, Germany\\
$^{3}$ Institute of Nuclear Research and Nuclear Energy\\
Bulgarian Academy of Sciences, Sofia 1784, Bulgaria\\}

\begin{document}
\maketitle 
\begin{abstract}
The information entropy of a nuclear density distribution is calculated for
a number of nuclei. Various phenomenological models for the density 
distribution using different geometry are employed. 
Nuclear densities calculated within various microscopic
mean field approaches are also employed. It turns out that the entropy 
increases on going from  crude phenomenological models to more 
sophisticated (microscopic) ones. It is concluded that the larger the information entropy, 
the better the quality of the nuclear density distribution.
An alternative approach is also examined: the net information content i.e.
the sum of information entropies in position and momentum space 
$S_{r}+S_{k}$. It is indicated that $S_{r}+S_{k}$ is a maximum, when the best
fit to experimental data of the density  and momentum  
distributions is attained.
\end{abstract}

\section{Introduction}

Information-theoretic methods have started recently to be important for the
study of quantum mechanical systems,
\cite{Ohya93,Bialynicki75,Aslangul72,Koga83,Gadre87,Gadre85,Gadre85A,
Majernik96,Panos97}
mainly through the application of the
maximum entropy principle (MEP) \cite{Kapur89,Canosa92}.
This is done by employing a suitably defined
quantum entropy that measures the lack of information associated with the
distribution of a quantum state over a given known basis. The MEP provides
the least biased description consistent with the available relevant 
information.

In \cite{Ghosh84} Ghosh, Berkowitz and Parr (GBP)
 defined, within the ground-state density
functional framework, several local quantities of thermodynamic-like character
in order to characterize a non-homogenous many-electron system.
Specifically, they introduced the concept of a local temperature corresponding
to the electronic motion and an information entropy $\cal S$ associated with
the electron distribution of an atomic system. An interesting
result \cite{Nagy96}
is that $\cal S$ increases with the quality of the wavefunction.

Gadre and Bendale (GB) \cite{Gadre87,Gadre85} investigated a
different MEP  procedure for atomic 
systems. They studied basis-set effects on the entropy sum ${\cal S}_{r} 
+ {\cal S}_{k}$, where ${\cal S}_{r}$ is the position-space entropy 
and ${\cal S}_{k}$ is the momentum-space
entropy. They found that this sum increases with the basis-set quality, so 
that ${\cal S}_{r} + {\cal S}_{k}$ can be used as a measure of the quality 
of the electronic density distribution.

A recent attempt to apply information theory to nuclear systems
is ref. \cite{Panos97}  
where the sum ${\cal S}_{r} + {\cal S}_{k}$ has been calculated employing 
two phenomenological models for the nucleus.
An interesting result was that the same functional form $ {\cal S} = 
aN + bNlnN$ for the entropy as function of the particle number 
N (electrons or nucleons) holds approximately for atomic and nuclear systems. 

The nucleon distributions in nuclei is of fundamental importance in our 
understanding of nuclear properties. There are several functional forms
which are used for the description of the nuclear distributions. 
Their parameters are usually fitted to the available empirical data for
the mean square  radii.
Moreover, density dependent Hartree-Fock (DDHF) \cite{Quentin78}
 calculations with 
Skyrme or Gogny type interactions or the relativistic mean field
(RMF) \cite{Ring96}
calculations  yield fairly reliable ground state nuclear 
properties including the densities. 

In the present paper we attempt to apply the approaches of GBP and
GB as a criterion of the quality of nuclear density distributions derived
according to various nuclear models. 
We hope that this will lead to a method to choose the best nuclear density
in different situations of practical interest, although as the models increase
in sophistication, it turns out that the value of $\cal S$ stops to increase 
and tends to saturation.

The paper is organized as follows: In section 2 we present the 
formalism of the information entropy according to GBP . In section 3 the
numerical results are shown and discussed. In section 4 an alternative 
approach (GB) i.e. the sum ${\cal S}_{r} + {\cal S}_{k}$ is discussed.
In section 5 numerical results for GB are presented and finally section
6 summarises our conclusions.

\section {The formalism of information entropy in phase-space}

In \cite{Ghosh84,Nagy96} Ghosh, Berkowitz and Parr  introduced a phase-space distribution
$f(\bar r,\bar p)$ associated with the ground-state density $\rho(\bar r)$ 
of a N-electron system. The distribution $f(\bar r,\bar p)$ satisfies 
the relations
\begin{eqnarray}
\rho(\bar r)&= &\int d\bar p f(\bar r,\bar p) \nonumber\\
\ \int d \bar r \rho(\bar r)& = &N \\
\ t(\bar r;\rho)& = &{1 \over 2} \int d \bar p p^2 f(\bar r,\bar p) \nonumber 
\end{eqnarray}
where $t(\bar r;\rho)$ is the kinetic energy density.

Then, they defined an entropy density $s(\bar r)$ and entropy $\cal S$ 
associated 
with the electron density $\rho(\bar r)$ in terms of $f(\bar r,\bar p)$ 
\begin{equation}
s(\bar r) = -k \int d \bar p f(lnf-1)
\end{equation}
\begin{equation}
{\cal S} = \int d \bar r s(\bar r)
\end{equation}
and obtained the most probable distribution $f(\bar r,\bar p)$ by maximizing 
this functional subject to the constraints of correct density $\rho(\bar r)$ and 
correct kinetic energy density $t(\bar r;\rho)$. 

The phase-space distribution function $f(\bar r,\bar p)$ is given by a local 
Maxwell-Boltzmann distribution law: 
\begin{equation}
f(\bar r,\bar p) = (\beta(\bar r) / 2\pi)^{3/2} \rho(\bar r) 
exp(-{1 \over 2}p^2\beta( \bar r))
\end{equation} 
where $\beta(\bar r)= 1/kT(\bar r)$. The local temperature $T(\bar r)$ is 
defined by the 
ideal-gas expression for the kinetic energy: 
\begin{equation}
t(\bar r;\rho) = (3/2) \rho(\bar r)kT(\bar r)
\end{equation}
Substituting eq. (4) into eq. (3) yields the Sackur-Tetrode
equation \cite{Ghosh84}  for
the entropy:
\begin{equation}
{\cal S} = k \int \rho [ {5 \over 2} -ln\rho + {3 \over 2} ln(2\pi kT)]dr
\end{equation}
The entropy density can also be rewritten in the form:
\begin{equation}
s(r) = {3 \over 2} k\rho[ln(t/t_{TF}) + c] 
\end{equation} 
where $t_{TF} = c_{k}\rho^{5/3}$ is the Thomas-Fermi kinetic energy density.
For convenience, $\bar r$ and $\rho$ are dropped from the
argument of $t$.
The constants $c$ and $c_{k}$ are given by:
\begin{equation}
c= {5 \over 3} + ln (4\pi c_{k}/3), ~~~~~~~~~~~ c_{k} = {3\over 10} (3\pi^2)^{2/3}
\end{equation}

In \cite{Nagy96} it was found that the entropy increases with the quality of the 
electron wavefunction. As stated in \cite{Ghosh84}, one can omit
all considerations 
for $f(\bar r,\bar p)$ and simply postulate the existence of (7)
and (3). 
Thus we postulate these expressions for another many fermion system i.e. the nucleus.
In the present article, we attempt to answer the question: Does this $\cal S$
increase with the quality of nuclear density distribution as well?
It turns out that the answer is yes. In other words, we derive a criterion
to assess the quality of a nuclear density distribution $\rho(r)$ by observing
an increase of $\cal S$ on going from crude to more sophisticated models of 
the nucleus.

\section{Numerical results for $\cal S$}

It is well known that the kinetic energy density $t$ is not uniquely defined.
Thus, there are various expressions for this quantity
\cite{Nagy96}: 
\begin{equation}
t_{1} =  {1 \over 8} \sum_{i} {{|\nabla\rho_i|^2} \over {\rho_i}} -{1\over 4}
\nabla^2 \rho 
\end{equation}
\begin{equation}
t_{2} = {1 \over 8} \sum_{i} {{|\nabla\rho_i|^2} \over {\rho_i}} \\ 
\end{equation}
\begin{equation}
t_{3} = {1 \over 8} \sum_{i} {{|\nabla\rho_i|^2} \over {\rho_i}} -{1\over 8}
\nabla^2 \rho
\end{equation}
  
For simplicity we have chosen the Thomas-Fermi-Weizs\"acker kinetic energy
expression \cite{Nagy96}:
\begin{equation}
t_{TFW} = t_{TF}+ {1 \over 72} {|\nabla \rho|^2 \over \rho} 
\end{equation}
This choice allows us to use  various functional forms of the density 
distributions, where no knowledge of the contribution to the density from each 
orbit is necessary. This is also corroborated
indirectly by the fact that in \cite{Nagy96}, using (12), there is not a single exception to the
trend of the information entropy to increase as the quality of the electron
wavefunction increases.

We calculate $\cal S$ employing as an input several nuclear (matter) density
distributions: 

1) Uniform (UN) specified by the well-known rule: $R = r_{0}A^{-1/3}$
\begin{equation}
\rho(r) = \left\{\begin{array}{cc}
          \rho_{0}   & \mbox{$r \leq R$} \\ 
           0         & \mbox{$r > R$}
           \end{array}
           \right.
\end{equation}

2) Trapezoidal (TR) \cite{Grypeos87}: 
\begin{equation}
\rho(r) = \left\{ \begin{array}{ll}
           \rho_{0}           &\;\;      \mbox{ $r < c-z$} \\
           \rho_{0}(c+z-r)/2z &\;\; \mbox{$c-z \leq r \leq c+z$} \\
           0                  & \;\;   \mbox{$r > c+z$}
           \end{array}
           \right.
\end{equation}
where $c$ is the half-way radius and $2z$ is the width of the
surface region.

3) Two parameter Fermi (FM) \cite{Daskaloyannis83}: 
\begin{equation}
 \rho (r) = {\rho_{0} \over {1+exp\left({{r-R}\over \alpha}\right)}} 
\end{equation}  
For the half density radius R, instead of the well known 
expression $R=r_{0}A^{-1/3}$ an expression with 
an A-dependent radius parameter $r_{0}= r_{0}(A)$ is used, which results
from the solution of an algerbraic third order equation for the 
normalization of the density \cite{Daskaloyannis83}:  

\begin{eqnarray}
r_{0}(A) = ~~~~~~~~~~~~~~~~~~~~~~~~~~~~~~~~~~~~~~~~~~~~~~~~~~~~~~~
~~~~~~~~~~~~~~~~~~~~~~~~~~~~~~~~~~~~~~~~ \nonumber \\ 
\left( {1 \over{2^{1/3}}} \right) r_{0}
\left\{ \left [ 1 + \left [1+{4 \over 27}
\left({\pi \alpha \over {r_{0}A^{1/3}}}\right)^{6}\right]^{1/2}\right]^{1/3}
+\left [ 1 - \left [1+ {4 \over 27}
\left({\pi \alpha \over {r_{0}A^{1/3}}}\right)^{6}\right ]^{1/2}\right ]^{1/3}
\right\}
\end{eqnarray}
where
\begin{equation}
 r_{0} = \left ( 3/ 4 \pi \rho_{0} \right )^{1/3} 
\end{equation}

It is noted that the same expression for the half density radius has also
been used for the trapezoidal distribution $(c = R)$, where $\pi \alpha=z$.
In both cases 
the parameters were determined by a global least squares procedure to the
available experimental rms radii.

4) Harmonic Oscillator (HO) Shell Model, specified by the rule : 
$\hbar \omega = 41 A^{-1/3}$

5) Semiphenomenological single-particle density (SP-D) of Gambhir and Patil
\cite{Gambhir85,Gambhir86,Gambhir89,Lalazissis97A}. We use the
expression of ref. \cite{Gambhir85}
\begin{equation}
\rho_{i}(r) ={ \rho^{0}_{i} \over 1+\beta_i[1 +
({r \over  {R+a_i}})^{2}]^{\alpha_{i}}
[e^{(r-R) \over {a_{i}}} +e^{-(r+R) \over {a_{i}}}]}
\end{equation}
\noindent
where
$$
\beta_i= [1 +({R \over  {R+a_i}})^{2}]^{-\alpha_{i}}
$$
\noindent 
(in order to idendify $R$ with the half-way radius), 
the index $i=n$ (for neutrons) or $p$ (for protons), $R$ is a
measure of the size of the nucleus and $a_{i}$ and $\alpha_{i}$
are given in terms of the separation energy $\epsilon_{i}$ of the 
last particle (neutron or proton) through:
\medskip
\begin{equation}
a_{i} ={ \hbar \over {2 \sqrt{2m\epsilon_{i}}}};
\end{equation}
\begin{equation}
 \alpha_{i} = 
{q \over \hbar} \sqrt{ m \over 2\epsilon_{i}} +1           
\end{equation}
\medskip
\noindent Here $m$ is the nucleon mass and  $q=0$ for neutrons and $q=Z-1$ 
for protons. 

The above $\rho_{i}$(r) correctly incorporates the following
two important physical requirements:\\
a) The small $r$ (r $\to$   0) behaviour which implies that the density 
contains only even powers of $r$. \\
b) The asymptotic behaviour 
\begin{equation}
\rho_{i}(r) \to  r^{-2\alpha_{i}}exp(-r/a_{i})~~~~~~   (r \to \infty)
\end{equation}

In the present work we employ the matter distribution
$\rho(r)=\rho_{n}(r)+\rho_{p}(r)$, where $\rho_{n}$(r) is
normalized to the number of neutrons and $\rho_{p}$(r) 
to the number of protons , so that $\rho$(r) is
normalized to the number of particles A (mass number). 

6) Densities calculated within the framework of microscopic mean field
approaches. Namely, densities obtained from Hartree-Fock calculations with
density dependent forces of Skyrme type  (DDHF) \cite{Quentin78} 
and the ones using the 
relativistic mean field (RMF) theory \cite{Ring96}. 

We have computed the information entropy $\cal S$ using the above densities.
Our results are shown in Table 1.
For the HO potential results are provided only up to $^{40}$Ca, since
this model is not expected to give realistic results for heavier nuclei.
The results for the Skyrme and RMF theories
have been obtained with the effective forces SkM* \cite{Brack85}
and NL3 \cite{Lalazissis97} respectively.
It is interesting to note, however, that 
using different Skyrme or RMF parametrizations, we have obtained quite similar
results i.e. the estimate of the entropy is independent of the used effective
force in either theory.

We observe in Table 1 that ${\cal S}$ increases from left to the right i.e.
on going from crude to more sophisticated (microscopic) models,
although this trend is less clear to the right of the table,
where the value of $\cal S$ tends to saturation.
Therefore, we can conclude that the larger $\cal S$, the better the quality
of $\rho(r)$. 
It is also seen that from the various functional forms used for the
description of the densities, the Fermi density shows the best behaviour 
leading to values for $\cal S$ quite similar to those of the microscopic
models. It is also observed that SP-D gives also large $\cal S$ values
i.e. it is a good density according to the present criterion. This is expected
because SP-D reproduces fairly well the experimental data.
Moreover, the $\cal S$ values of SP-D are quite similar to those of Fermi 
distribution and one could say that they both lead to $\rho(r)$
of similar quality. The interesting thing is that the experimental input used for the
determination of their parameters is quite different. In Fermi distribution
the parameters are treated as free parameters determined by a global fit
to experimental rms radii while in SP-D the parameters (not free) are
adjusted to the experimental separation energies of the last proton and
neutron of each nucleus. 

\section{An alternative approach}

Gadre and Bendale (GB) \cite{Gadre87,Gadre85} employed a different maximum entropy procedure
for atomic systems. They studied the entropy sum ${\cal S}_{r} + {\cal S}_{k}$
where 
\begin{eqnarray}
 {\cal S}_{r} = &- \int \rho(\bar r) ln\rho(\bar r) d\bar r \\ 
\nonumber \\
 {\cal S}_{k} = &- \int n(\bar k) lnn(\bar k) d\bar k 
\end{eqnarray}
${\cal S}_{r}$ is the position-space information entropy and ${\cal S}_{k}$
is the momentum-space information entropy. $\rho(\bar r)$ and $n(\bar k)$
are the single-particle densities in configuration and momentum space 
respectively.
They found that the sum ${\cal S}_{r} + {\cal S}_{k}$ increases with the
basis-set quality of the electron density $\rho(r)$. 

In \cite{Panos97} we have focused on the same sum ${\cal S}_{r} + {\cal S}_{k}$ but
for another many-fermion system i.e. the nucleus. We have employed
two models of the nucleus, first the
nuclear mean field was approximated by a HO potential and then the model was 
extended by introducing some sort of short-range correlations.
In both approaches we have found that ${\cal S}_{r} + {\cal S}_{k}$ is
independent of the single parameter of the model (the HO parameter) i.e. 
it is scale invariant and characterizes every nucleus. We have also found
that short-range correlations increase slightly the net information content
${\cal S}_{r} + {\cal S}_{k}$ of the nucleus. The relative increase in
${\cal S}_{r} + {\cal S}_{k}$ due to short range correlations ranges from
2.5\% for $^4$He to 1.2\% for $^{40}$Ca. 

In the present work we calculate the sum ${\cal S}_{r} + {\cal S}_{k}$
for the three doubly magic nuclei $^4$He, $^{16} $O and $^{40}$Ca using as 
input $\rho(r)$ and $n(k)$ from a phenomenological method
\cite{Antonov96}  based on
the natural orbital representation to construct the one-body density matrix
(OBDM) $\rho(\bar r, \bar r^{\prime })$. According to this method:
\begin{eqnarray}
\rho(\bar r)& \equiv &\rho(\bar r,\bar r) = \sum_{i} \lambda_{i}
|\psi_{i}(\bar r)|^{2}
\nonumber \\
\     \\
n(\bar k)& \equiv &n(\bar k,\bar k) = \sum_{i} \lambda_{i}
|\tilde \psi_{i}(\bar k)|^{2}
\nonumber
\end{eqnarray}  
where $i=nlm$, 
$\tilde \psi_{i}(\bar k)$ is the Fourier transform of
$\psi_{i}(\bar r)$ and 
$\lambda_{nl}$ are the occupation numbers. The natural orbitals can be 
looked for in the form:
\begin{equation}
\psi_{i}(\bar r) \equiv \psi_{nlm}(\bar r) = R_{nl}(r)Y_{lm}(\theta,\phi)
\end{equation}
The radial part of the natural orbitals of the OBDM 
are expanded \cite{Antonov96} in terms
of single-particle wavefunctions $\left \{ \phi_{nl}(r) \right \}$ preserving
all the usual symmetries for spherical nuclei:
\begin{equation}
R_{\alpha} = \sum^{3}_{i=1} C_{i}^{\alpha} \phi_{il}(r) ~~~ (\alpha \equiv nl)
\end{equation}
In ref. \cite{Antonov96} three sets of s.p. wavefunctions 
$\left \{ \phi_{nl}(r) \right \}$ were used corresponding to:

\noindent 1) the harmonic oscillator potential (HO)
\begin{equation}
V(r) = -V_{0} + {1 \over 2} m\omega^{2}r^{2} ~~~~(V_{0} > 0)
\end{equation} 
2) the square well (SW) potential with infite walls
\begin{equation}
V(r) = \left \{ \begin{array}{ll}
       -V_{0} &\;\; \mbox{ $r < x, \;\;\;V_{0} > 0$}\\
        \infty&\;\; \mbox{$ r > x$}
       \end{array}
       \right.
\end{equation}
3) the modified harmonic-oscillator (MHO) potential \cite{Ypsilantis} 
\begin{equation}
V(r) = -V_{0} + {1 \over 2} m\omega^{2}r^{2} + {B \over r^{2}}, ~~~~
V_{0} > 0 , B \ge 0
\end{equation}
The parameters of the above expressions were obtained in \cite{Antonov96} by fitting
simultaneously the local density $\rho(r)$ and the momentum distribution
n(k) to the available empirical data. However, while for the density
distribution a lot of information is available, this is not the case 
with the momentum distribution where the information is rather limited.
Experimental data for n(k) is available only for $^{4}$He \cite{Ciofi91}.
For this reason,
the theoretical estimates of n(k) in $^{16}$O and $^{40}$Ca obtained
within the Jastrow correlation method (JCM) \cite{Stoitsov93}
were used in the fit. 
In all cases examined the best-fit values were obtained 
using the square-well  single-particle wavefunctions.

\section{Numerical results for the sum ${\cal S}_{r} + {\cal S}_{k}$}

In table 2, we list for comparison the ${\cal S}_{r}$,
${\cal S}_{k}$ and ${\cal S}_{r} + {\cal S}_{k}$ values
calculated by inserting (24) into equations (22),(23)  
for each set of basis wavefunctions (HO, SW, MHO). 
It is observed that ${\cal S}_{r} + {\cal S}_{k}$ for
$^{4}$He is a maximum for the SW case, where also the best fit to the
experiment is obtained. However, this is not the case for 
$^{16}$O and $^{40}$Ca nuclei where ${\cal S}_{r} + {\cal S}_{k}$ becomes
maximum for the HO case. It should be noted, however, 
that for $^{16}$O and $^{40}$Ca theoretical input for n(k) has been 
used in the fitting procedure. Moreover the theoretical n(k) values obtained
within JCM were calculated using as only input experimental knowledge
of the density distributions. 
  
Our results suggest that when experimental 
data ($^{4}$He) are reproduced well, the sum  ${\cal S}_{r} + {\cal S}_{k}$
is indeed a maximum, i.e. the larger ${\cal S}_{r} + {\cal S}_{k}$, the
better the quality of $\rho(r)$.  
This also indicates that more experimental information for n(k) is called for.
It is also seen that in $^{4}$He 
where n(k) is deduced from experiment the ${\cal S}_{k}$ value is 
larger in SW case  than in the HO and MHO cases where the values
are smaller and very close to each other.

\section{Discussion and conclusions}
We have presented two methods of assessing the quality of nuclear density
distributions. The first method is based on the information entropy $\cal S$
associated with phase-space distribution. It is seen that the Fermi 
distributions and the densities deduced from microscopic models correspond 
to larger information entropies i.e. they are of better quality compared 
to more crude models of the nucleus.
It is also observed that the semiphenomenological density (SP-D) derived
by reproducing the separation energies of the last nucleon and the
correct asymptotic behaviour of $\rho(r)$ is of comparable quality
(according to the present criterion) with Skyrme DDHF and RMF approaches.
This is expected because SP-D reproduces well the empirical values for the
rms radii.
Another observation is that different parametrizations of Skyrme and RMF
models give similar results for the information entropy $\cal S$.
It seems that after a certain degree of sophistication of a nuclear model
is reached, the information entropy no longer increases but saturates.
However, such a conclusion needs more examination. 
  
The second method examines the sum $S_{r} + S_{k}$ of the entropies in
position and momentum space. We have used a certain phenomenological model
which constructs the one-body density matrix  $\rho(r,r^{'})$.
The parameters of the corresponding densities $\rho(r)$ and $n(k)$ were found
by fitting simultaneously the experimental data of $\rho(r)$ and $n(k)$. 
For $n(k)$, however, only for $^{4}$He some experimental input is available.
Therefore, the predictions of a model (JCM) were used as an input for the rest of
the cases ($^{16}$O and $^{40}$Ca).
Our study shows that for $^{4}$He and for wavefunctions 
corresponding to the best quality fit (SW), among those considered, 
the sum $S_{r} + S_{k}$ gets its highest value. 
Moreover, in $^{4}$He, where experimental 
input for n(k) was used, the  $S_{k}$ value for the best quality 
wavefunctions is significantly larger compared with the values obtained with
the use of the other wavefunctions. 
For $^{16}$O and $^{40}$Ca , however, the sum becomes maximum for HO 
wavefunctions which do not correspond to the best fit.  This should be 
connected with the use of theoretical predictions for $n(k)$ due to lack of
experimental information. These results, however, should be considered 
preliminary because a limited number of cases have been examined in the 
OBDM method of ref. \cite{Antonov96}, though they are in the correct direction and 
suggest that the MEP works, in accord with the first method of the
present paper.

\begin{table}[h]
\caption{ \sf The information entropy $\cal S$ for several
nuclei using various nuclear distributions (GBP approach)} 
\begin{center}
\begin{tabular}{|l  c c c c c c c c | }
\hline
\ Nucleus&UN & TR & HO  & SP-D & FM  & DDHF & RMF & \\
\hline\hline
$^{16}$~O&~99.69&101.68&102.06&102.11&102.12&102.10&102.13& \\
$^{32}$~S&199.38&202.34&202.75&202.81&202.88&202.80&202.90& \\
$^{40}$Ca&249.23&252.59&253.04&253.08&253.18&253.11&253.17& \\
$^{90}$Zr&560.77&566.10&      &566.66&566.96&566.90&566.94& \\
$^{116}$Sn&722.76&728.68&     &729.45&729.91&729.97&729.88& \\
$^{208}$Pb&1295.99&1304.57&   &1304.91&1305.99&1305.81&1306.20& \\
\hline
\end{tabular}
\end{center}
 \end{table}


\begin{table}[h]
\caption{ \sf The information entropies ${\cal S}_{r}$, ${\cal
S}_{k}$ and ${\cal S}_{r}+{\cal S}_{k}$ for three nuclei using
three sets of single particle wavefunctions corresponding to
harmonic oscillator (HO), square well(SW) and modified harmonic
oscillator (MHO) potentials (GB approach)} 
\begin{center}
\begin{tabular}{|l c c  c c c c c c c c | }
\hline
\ Nucleus& &  HO&  &  & SW& & & MHO & & \\
\hline
        & ${\cal S}_{r}$ & ${\cal S}_{k}$ & ${\cal S}_{r} + {\cal S}_{k}$&
          ${\cal S}_{r}$ & ${\cal S}_{k}$ & ${\cal S}_{r} + {\cal S}_{k}$&
          ${\cal S}_{r}$ & ${\cal S}_{k}$ & ${\cal S}_{r} + {\cal S}_{k}$&\\
\hline
$^{~4}$He&7.27&2.06&9.33&6.91&4.26&11.17&7.10&2.28&9.38&\\
$^{16}$~O&27.96&3.69&31.65&27.91&2.80&30.71&28.02&1.20&29.22&\\
$^{40}$Ca&66.22&7.82&74.04&65.67&-5.81&59.86&66.38&-9.09&57.29&\\
\hline
\end{tabular}
\end{center}
\end{table}


\newpage

\begin{thebibliography}{99}
\bibitem{Ohya93} M. Ohya and D. Petz, "Quantum entropy and its
use", Springer-Verlag (1993) 
\bibitem{Bialynicki75} I. Bialynicki-Birula and J. Mycielski,
Commun. Math. Phys. {\bf 44} 129 (1975).
\bibitem{Aslangul72} C. Aslangul, R. Constanciel, R. Daudel and
P. Kottis, Adv. Quantum Chem. {\bf 6} 94 (1972). 
\bibitem{Koga83} T. Koga and M. Morita, J. Chem. Phys. {\bf 79}
1933 (1983).
\bibitem{Gadre87} S. R. Gadre and R. D. Bendale, Phys. Rev. A
{\bf 36} 1932 (1987).
\bibitem{Gadre85} S. R. Gadre and R. D. Bendale, Current Sci.
{\bf 5} 970 (1985).
\bibitem{Gadre85A} S. R. Gadre, S. B. Sears, S. J. Chakravorty
and R. D. Bendale, Phys. Rev. A {\bf 32} 2602 (1985)
\bibitem{Majernik96} V. Majernik and T. Opatrny, J. Phys. A :
Math. Gen. {\bf 29} 2187 (1996).
\bibitem{Panos97} C. P. Panos and S. E. Massen, Int. J. Mod.
Phys. E {\bf 6} 497 (1997)
\bibitem{Kapur89} J. N. Kapur " Maximum-Entropy Models in
Science and Engineering" John Wiley \& Sons, Inc. (1989) 
\bibitem{Canosa92} N. Canosa , A. Plastino and R. Rossignoli,
Nucl. Phys. A {\bf 550} 453 (1992).
\bibitem{Ghosh84} S.K. Ghosh, M. Berkowitz and R. G. Parr, Proc.
Natl. Acad. Sci. USA {\bf 81} 8028 (1984)
\bibitem{Nagy96} A. Nagy and R. G. Parr, Int. J. Quant. Chem.
{\bf 58} 323 (1996)
\bibitem{Quentin78} P. Quentin and H. Flocard, Ann. Rev. Nucl.
Sci. {\bf 28} 523 (1978)
\bibitem{Ring96} P. Ring, Prog. Part. Nucl. Phys. {\bf 37} 193
(1996) 
\bibitem{Grypeos87} M. E. Grypeos and C. P. Panos, Int. J. Theor.
Phys. {\bf 26} 583 (1987)
\bibitem{Daskaloyannis83} C. B. Daskaloyannis, M. E. Grypeos, C.
G. Koutroulos, S. E. Massen and D. S. Saloupis, Phys. Lett. {\bf
121B} 91 (1983)   
\bibitem{Gambhir85} Y. K. Gambhir and S. H. Patil, Z. Phys. A
{\bf 321} 161 (1985)
\bibitem{Gambhir86} Y. K. Gambhir and S. H. Patil, Z. Phys. A
{\bf 324} 9 (1986)
\bibitem{Gambhir89} Y. K. Gambhir, P. Ring and H. De Vries,
Europhys. Lett. {\bf 10} 219 (1989)
\bibitem{Lalazissis97A} G. A. Lalazissis, C. P. Panos, M. E.
Grypeos and Y. K. Gambhir, Z. Phys. A {\bf 357} 429 (1997)
\bibitem{Brack85} M. Brack, C. Guet and H. B. Hakanson, Phys.
Rep. {\bf123} 275 (1985).
\bibitem{Lalazissis97} G. A. Lalazissis, J. K\"onig and P. Ring,
Phys. Rev. C {\bf 55} 540 (1997)
\bibitem{Antonov96} A. N. Antonov, S. S. Dimitrova, M. K.
Gaidarov, M. V. Stoitsov, M. E. Grypeos, S. E. Massen and K. N.
Ypsilantis, Nucl. Phys. A {\bf 597} 163 (1996).
\bibitem{Ypsilantis} K. Ypsilantis and M. Grypeos, Nuovo Cim.
{\bf 82} A 93 (1984); M. E. Grypeos and K. Ypsilantis, J. Phys.
G {\bf 15} 1397 (1989); K. Ypsilantis and M. E. Grypeos, J.
Phys. G {\bf 21} 1701 (1995).
\bibitem{Ciofi91} C. Ciofi degli Atti, E. Pace and G. Salme,
Phys. Rev. C {\bf 43} 1155 (1991)
\bibitem{Stoitsov93} M. V. Stoitsov, A. N. Antonov and S. S.
Dimitrova, Z. Phys. A {\bf 345} 359 (1993)
 
\end{thebibliography}
\end{document}